\documentclass{sig-alternate-10pt}

\usepackage{times}  

\usepackage[small]{caption}
\usepackage{graphicx}
\usepackage{amssymb}
\usepackage{amsmath}
\usepackage{epstopdf}
\usepackage{epsfig}
\usepackage{float}
\usepackage{algorithmic}
\usepackage{enumerate}
\usepackage{xspace}
\usepackage{comment}
\usepackage{color}
\usepackage[usenames,dvipsnames,svgnames,table]{xcolor}
\usepackage{placeins}
\usepackage{tabularx, lipsum}
\usepackage{multirow}
\usepackage{enumitem}

\usepackage[paperwidth=8.5in, paperheight=11in, hmargin=0.75in, vmargin=0.875in, textwidth=7in,
columnsep=0.33in]{geometry}

\newenvironment{Itemize}%
{\vspace{-1.3mm}
\begin{itemize}[leftmargin=0.5cm]
\setlength{\itemsep}{0pt}%
\setlength{\topsep}{0pt}%
\setlength{\partopsep}{0pt}%
\setlength{\parskip}{0pt}}%
{\end{itemize}
\vspace{-1.3mm}}

\newenvironment{Enumerate}%
{\vspace{-1.3 mm}
\begin{enumerate}[leftmargin=0.5cm]%
\setlength{\itemsep}{0pt}%
\setlength{\topsep}{0pt}%
\setlength{\partopsep}{0pt}%
\setlength{\parskip}{0pt}}%
{\end{enumerate}
\vspace{-1.3 mm}} 

\newcommand{\denselist}{\itemsep -2pt\parsep=1pt\partopsep 0pt}
\newcommand{\bitem}{\begin{itemize}\denselist}
\newcommand{\eitem}{\end{itemize}}
\newcommand{\benum}{\begin{enumerate}\denselist}
\newcommand{\eenum}{\end{enumerate}}

\DeclareMathOperator{\Ex}{E}  

\newcommand{\name}{SwiftC\xspace}
\newcommand{\tit}{\textit}
\newcommand{\tbf}{\textbf}
\renewcommand{\vec}[1]{\mathbf{#1}}

\newcommand{\squeezeup}{\vspace{-2.5mm}}
  
\begin{document}
\title{A low-latency control plane for dense cellular networks}
\author{Rakesh Misra, Sachin Katti\\
		\affaddr{Stanford University}\\
		\affaddr{\{rakeshmisra, skatti\}@stanford.edu}}
\maketitle

\pagestyle{plain}
\pagenumbering{arabic}

\section*{Abstract}

In order to keep up with the increasing demands for capacity, cellular networks
are becoming increasingly dense and heterogeneous. Dense deployments are
expected to provide a linear capacity scaling with the number of small cells
deployed due to spatial reuse gains. However in practice network capacity is
severely limited in dense networks due to interference. The primary reason is
that the current LTE control plane deployment model has very high latency and is
unable to cope with the demand of implementing interference management
techniques that require coordination on a millisecond timeframe. 

This paper presents \name, a novel low-latency control plane design for LTE
networks. \name's novel contribution is a design for efficiently sending and receiving
control plane messages over the LTE spectrum itself, thus creating a direct and
low-latency coordination signaling link between small cells and the macro
cell. \name builds on recent work in full duplex radios and shows via prototype
implementations that a low latency control plane can be built over the existing
LTE network without wasting licensed spectrum. We also show the benefits of
\name in implementing complex interference management techniques, and show that
with \name small cell deployments can achieve almost a linear capacity scaling
with every small cell deployed.  

\section{Introduction}

Dense cellular networks with several small cells deployed per macrocell are
expected to provide significant gains, increasing capacity almost linearly with
every deployed small cell. The benefits come from two factors: clients are
closer to the base station (BS) leading to higher SNR links, and the load per BS
is significantly lower due to the smaller coverage area leading to higher per
user throughput and higher spatial reuse. Consequently, there is tremendous
interest from industry to move towards such
architectures~\cite{bib:metrocell},~\cite{bib:nanogsm} and preliminary
deployments have already begun.

A big challenge however is the lack of licensed spectrum. Ideally, operators
would like to operate adjacent small cells on different frequencies so as to
avoid interference and get the capacity scaling benefits.  Unfortunately,
licensed spectrum is scarce and expensive.  For example, Verizon and AT\&T own
only 20MHz of contiguous downlink spectrum each licensed for LTE in the 700MHz
band, which is too small to divide into multiple channels efficiently (dividing
requires guard bands which can be quite wasteful given the limited spectrum
widths).  Consequently, adjacent small cells are expected to operate on the same
channel.  In industry parlance, this is referred to as network deployments with
a spectrum reuse factor of one, which is how current LTE networks are being
deployed.

The problem therefore is that small cells are going to cause significant
interference to each other as well as to the macro BS whose larger coverage area
spans the small cell coverage areas.  Further, since licensed spectrum is
typically in the lower carrier frequencies (700-900MHz or 1.8GHz), radio signals
propagate farther, and thus interference is exacerbated.  Left unmanaged,
interference will severely limit network capacity and prevent small cells from
achieving the expected linear capacity scaling. 

To cope, the cellular standards body 3GPP has introduced several control plane
mechanisms to manage interference. The default mode is to just ignore the
interference  (i.e., treat it as noise) and this is in fact the most widely used
mechanism today, but as we will show in Section~\ref{sec:motivation}, in this
case network capacity is severely limited by interference.  Recently, a newer
set of mechanisms known as eICIC (enhanced Inter Cell Interference Coordination)
has been introduced which essentially tries to schedule transmissions from the
macro BS and the small cells in different time-frequency slots so that they
avoid interfering with each other.  A final more sophisticated mechanism is CoMP
(Coordinated Multi Point), which attempts to exploit interference by jointly
coding the transmission across multiple small cell BSes such that these BSes
mimic a large MIMO antenna array.  This mechanism requires the exchange of
channel state information between the BSes, coordination on a joint pre-coding
vector at all the BSes and synchronized transmission.  This mechanism can
provide large benefits, and unlike the other two mechanisms, can in fact turn
interference into an advantage rather than a liability 

Our problem thesis in this paper is that the current control plane and
deployment architecture for the radio access network (RAN) in cellular networks
is not capable of realizing the interference coordination mechanisms described
above. The key challenge is latency. The LTE control plane is based on the X2
interface~\cite{bib:3gppx2} which is a logical interface defined over the
physical backhaul links used to connect the BSes back to the packet core of the
operator.  The physical backhaul links use a variety of technologies, mostly
out of necessity.  Given the density of small cell deployments, and the fact
that they have to be deployed in urban areas, operators are forced to use
whatever backhaul connectivity is available (ranging from cable networks to DSL
to public fiber if its available).  All of these physical network options
exhibit latencies of 10-15ms to the cellular packet core in the best case, and
often much higher.

Such high latencies pose problems in managing interference.  Specifically, our
analysis in Section~\ref{sec:motivation} shows that with such latencies, network
performance can often be worse with interference management mechanisms such
as CoMP than without!  The reason is that these mechanisms depend upon
neighboring BSes coordinating with each other by exchanging channel state
information and making joint scheduling and coding decisions. However such
coordination has to occur at the same granularity as channel coherence times,
which is around $4-5$ms (90\% coherence) even for static and walking mobility users in urban
environments. Hence a coordination latency of 10ms would mean that the channel state
information that was shared is only about 80\% correlated with the actual
channel during transmission (or in other words, has a \textit{staleness} of
20\%, as we define later). Staleness shows up as increased interference and
decrease in SINR, reducing network capacity. The effect is worsened in dense
networks because the increase in interference from staleness is proportional to
the channel strength, which is by definition high in dense networks.

To tackle this problem, in this paper we present the design and implementation
of a novel low-latency and efficient control plane, \name, for cellular
networks. \name's key architectural insight is to \emph{physically} decouple the
control and data plane backhaul connectivity and use different connectivity
options for them. We argue that imposing the requirement that a single backhaul
technology satisfy the stringent latency requirement that the LTE control plane
requires and also the high bandwidth requirement that the data plane desires
leads to unnecessarily expensive and complex backhaul designs, which in turn
makes small cells expensive or infeasible to deploy. We argue that its in fact
better to design the backhaul for the control plane separately to satisfy the
low latency requirement. 

\name's physical control plane design is built on the LTE air interface itself as the
physical technology, in other words we use the macro LTE network itself as the
control plane backhaul network for the small cell.  Consequently \name inherits
the latency guarantees that LTE itself provides, namely 1ms best case latency if
needed for urgent transmissions (the size of the LTE subframe).  The natural
concern with this approach of course is whether we are using up scarce LTE
spectrum to build a control plane.  In this paper we present a novel design for
the small cells that allows us to build this control plane almost for free by
leveraging recent work on full duplex radios.  Specifically, we
enable each small cell to connect to the macro base station as if it is a phone,
on the same frequencies it is using for serving clients connected to itself.  The idea
is that when the small cell is receiving transmissions from the phones connected
to it on the uplink frequency, it can turn around and transmit control frames to
the macro base station at the same time on the same frequency because of full
duplex capability.  The idea works analogously in the downlink direction, at the
same time as the small cell is receiving control messages from the macro BS on
the downlink frequency, it can transmit to the phones connected to it.  Hence no
capacity is lost at the small cell, and the control is piggybacked on the same
frequencies. Meanwhile the data plane backhaul can be provided using whatever
connectivity is available as long as it has sufficient bandwidth, it does not
have to also satisfy the low latency requirement at the same time. 

We design and implement \name and show the benefits of such a low latency 
control plane in managing interference in dense cellular networks. 
We show that with \name, overall network capacity can scale almost linearly 
with the number of small cells deployed. 

Finally, we note that recently some operators such as China Mobile have proposed
C-RAN~\cite{bib:cran}, an architecture where the entire baseband of multiple base stations is
centralized and processed in a datacenter. The idea is that each base station
would only be a simple radio, and they would send/receive IQ digital samples
itself from the centralized datacenter. So all the signal processing and protocol
handling is done completely in the centralized location, and consequently C-RAN
can definitely implement all of the above interference control mechanisms
easily. However C-RAN is extremely expensive and in many scenarios impossible to
deploy. This is because C-RAN requires that every base station be connected by
direct fiber links with less than 5ms latency with several gigabits of
bandwidth. Such a fiber deployment is extremely expensive, and in many cases
infeasible in Europe and the US because cities have already been built out.
\name is also architecturally simpler. Operators now have the flexibility to
chose whatever backhaul is available for data backhaul without having to worry
about whether that option can satisfy the stringent requirements for the RAN
control plane. We believe this can greatly simplify small cell deployment.

\section{Background}

\subsection{LTE Small Cells and the Control Plane}
The 3GPP LTE standards define two base station types (also called evolved NodeBs or 
eNBs) that are meant for small cell deployments~\cite{bib:3gppbs} - medium range eNBs 
(metro or micro cells) and local area eNBs (pico cells), see Table~\ref{tab:small_cell}. 
Future urban outdoor networks are expected to be increasingly heterogeneous 
with these small cells deployed within existing macro cell coverage areas to increase 
network capacity and extend service coverage. 
Typical deployment models suggest deploying around 4-12 small cells per macro sector, 
or 12-36 small cells per macro eNB assuming a 3-sector macro eNB.
With increasing density, there is going to be an increasing need for coordination between 
groups of small cells as well as between small cells and macrocells in order to use the 
radio resources more efficiently.
\begin{table}[h]
	\centering
	\caption{Typical urban small cell properties}
	\begin{tabular}{|c|c|c|}
	\hline
   					&	\tbf{Microcell}		& 	\tbf{Picocell} 	\\ \hline
	\tbf{Max Tx power}	&	+38dBm			&	+24dBm		\\ 
	\tbf{Range}		&	200-500m			&	50-100m		\\ 
	\tbf{Location}		&	Outdoor			&	Indoor/Outdoor	 \\ 
	\tbf{Simul. Active Users}	& 	32-64			&	32-64 \\ \hline	
	\end{tabular}
    	\label{tab:small_cell}
\end{table}
 
The 3GPP LTE standards specify a control plane for LTE to implement interference
coordination and other network management functions. 
The control plane is built on an X2 interface that interconnects neighboring base
stations~\cite{bib:3gppx2}.  
X2 is a logical point-to-point interface that can be switched over the existing IP 
transport network that is serving as the backhaul for the base station to the operator's 
packet core network. 
The typical organization of the control plane has the macro eNB acting as the logically
centralized point with network and user state information being fed via the X2
interface to this coordinator. 
However, in many cases, the X2 interface is directly used to connect two neighboring 
small cells when they need to make a localized control decision such as handovers. 
So the LTE control plane is hybrid; distributed for certain functions (handovers) and
logically centralized for some others (interference, load management etc). 
In this paper we will focus on the interference management aspects of the control plane
design since that is typically the most important factor that determines overall
network performance and spectral efficiency.

The latency and bandwidth of the control plane is therefore dictated primarily
by the quality of the backhaul link technology and network used to connect the
BSes to the packet core network. While operators typically invest significantly
in providing macro BS with dedicated fiber backhaul, they cannot do so for small
cells because of their smaller coverage area. The economics of a small cell (in
terms of its coverage area and the number of users it can support at any time)
does not justify deploying dedicated fiber links to each cell. Further small
cells are often deployed in urban hotspots (e.g. bus stops, concerts, downtown
etc) where digging and deploying fiber is prohibitively expensive or not
possible due to city regulations. Hence operators are forced to use any
available backhaul including cable connections, DSL lines etc that are leased
from another network operator such as Comcast. The IP latency of these links
from any small cell BS to a neighboring BS or to the macro coordinator is on the
order of 20-30ms. Bandwidth is typically not an issue since these links are
capable of providing 100Mbps speeds which is sufficient for LTE small cells.

Some companies have recently proposed using microwave frequencies (5-10GHz) for
building point to point backhaul links.  However typically these links do not
work well for small cell deployments in urban areas.  Most city areas where one
might deploy a small cell are likely to be concrete jungles, with no direct line
of sight path available between neighboring small cells or to the macro eNB.
Without LOS, links built on these higher frequencies do not work well since
signals at these frequencies do not travel through walls and lose a lot of
strength upon reflection. 

\subsection{Interference Management in the LTE Control Plane}

While deploying small cells is becoming increasingly necessary to keep up with
the ever-increasing demand for capacity, more small cells also means more
cell-edges in the network and therefore more problems of inter-cell interference
given that LTE is designed for a frequency reuse of 1.  LTE and LTE-Advanced
(LTE-A) specify several schemes for interference management, each requiring
varying degrees of coordination between neighboring base stations and thereby
having different latency and bandwidth requirements over the control plane.

\tit{Inter-Cell Interference Coordination} (ICIC) is specified in Releases 8/9
and involves frequency and power domain coordination whereby certain resource
blocks are avoided by one cell (or used to serve core users at lower powers) so
that they can be used by the edge users of neighboring cells.  ICIC requires
semi-static coordination on the order of a few seconds and is not sensitive to
control plane latencies that is on the order of tens of milliseconds, and also
has negligible bandwidth requirements compared to, for example,
handover~\cite{bib:x2}.  Recently, ICIC has been extended to time domain in
\tit{enhanced Inter-Cell Interference Coordination} (eICIC) wherein the
macrocell stops using the traffic channel in \tit{Almost Blank Subframes [ABS]}
(but keeps broadcasting essential signaling and information at very low power),
thereby allowing small cells to serve those UEs who would have otherwise
experienced strong interference from the macrocell.  Depending on the data
traffic demand, the ABS pattern needs to be coordinated on the order of
40ms~\cite{bib:ofdmabook}.
	
In Release 9, the semi-static coordination in ICIC is enhanced to dynamic
coordination in Coordinated Multi-Point (CoMP) where resources can be
coordinated as rapidly as on a per 1ms sub-frame basis.  One form of CoMP is
\tit{Coordinated Scheduling/Beamforming} (CS/CB) where neighboring cells
coordinate user scheduling and beamforming decisions with each other.  The more
attractive form of CoMP is \tit{Joint Transmission} (JT), a sub-category of
Joint Processing where data to a single UE is transmitted from multiple cells to
improve the received signal quality and/or actively cancel interference for
other UEs.  To do so, the cells have to exchange channel state information (CSI)
from each cell to each UE for which transmissions are being coordinated, and
then calculate the precoding vectors to use.  CoMP requires significantly more
backhaul bandwidth - one estimate suggests 770kbps of control information per X2
interface between tri-cell base stations coordinating every 1ms in CS/CB, and
several Mbps more of forwarded user data in JT~\cite{bib:qcx2} - as well as a
lower control plane latency on the order of a millisecond, and also requires
tight synchronization between the coordinating base stations.

It is important to mention here that while ICIC, eICIC and CS/CB essentially try
to \tit{avoid} or reduce interference, CoMP-JT tries to \tit{exploit} the
interfering links to provide capacity gains over other schemes. So ideally,
operators want to use CoMP-JT and achieve the higher capacity gains. 
In the next section, we discuss  the impact of the practical control plane
latencies on the performance of interference management and specifically CoMP\footnote{In
the rest of the paper, by CoMP we will mean CoMP-JT.} to motivate the need for a
low-latency control plane.

.

\section{Is the current control plane sufficient for LTE?}
\label{sec:motivation}

In this section, we show via detailed simulations and analysis that the current
LTE control plane design is incapable of scaling to meet the control signaling
needs for dense small cell heterogeneous networks. As described in the previous
section, in current deployments, the control plane latency is on the order of
tens of milliseconds, whereas bandwidth is plentiful (around 100Mbps or more).
We show that with such latencies, control plane functionality such as CoMP for
managing interference would work poorly and in fact hurt the performance of the
network.  Intuitively, the reason is the coherence time of the channels.  An
empirical estimate of the 90\% coherence time for an urban outdoor user walking
at 5kmph is 4-5ms~\cite{bib:rappaport}, a fact that is corroborated
by~\cite{bib:matlab}.  Thus, a 10ms latency implies that the shared channel
state information (CSI) is only about 80\% correlated with the true channel by
the time it is transported over the X2 control interface. We find that such
\tit{staleness} in CSI causes the average network capacity with CoMP to drop by 
as much as 40\% or more and is in fact worse than doing nothing (i.e. ignoring
interference) for core users and avoiding interference (i.e. by time-division) for
edge users. We briefly present our findings below.

\subsection{Impact of control plane latency}
\begin{figure}[htpb]
  \begin{center}
    \includegraphics[scale = 0.28]{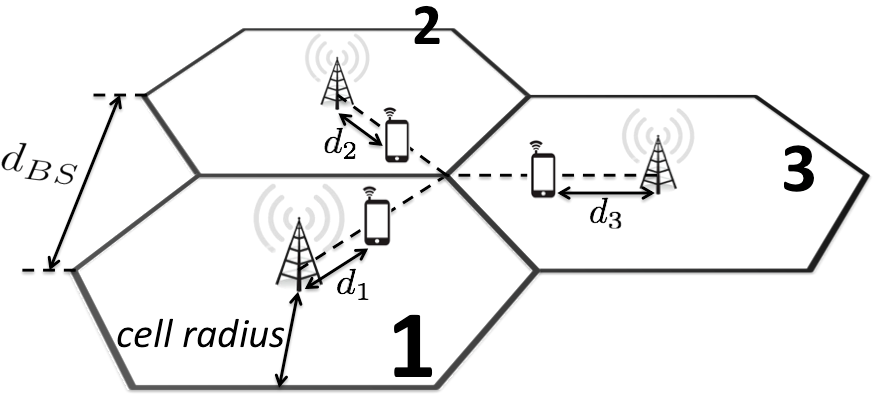}
    \caption{A $3\times 3$ urban microcell cluster}
    \label{fig:sim_topo}
  \end{center}
\end{figure}

\begin{figure}[htpb]
  \begin{center}
    \includegraphics[scale = 0.23]{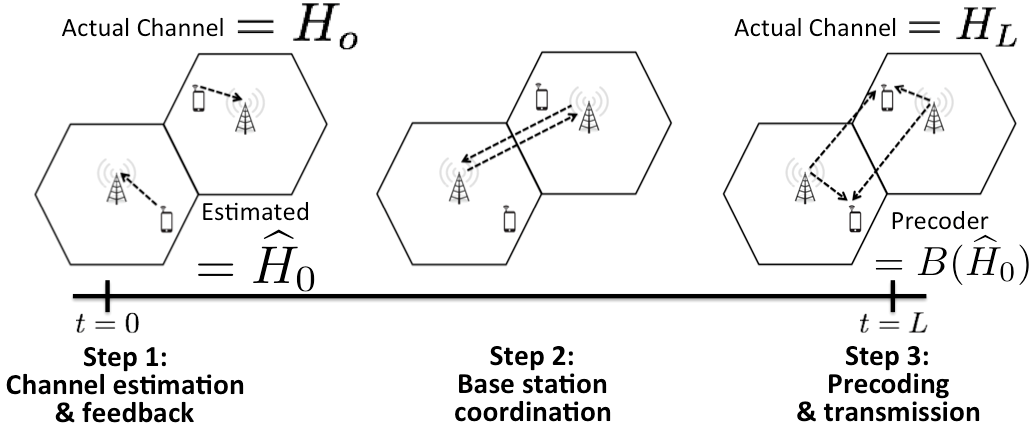}
    \caption{CoMP Joint Transmission. In order to focus on the impact of coordination latency
    alone, we do not incorporate any channel estimation errors here, so $\widehat{H}_0 = H$.
    The soft interference nulling precoder $B$ is computed from a stale version $H_0$ of the
    channel when the true channel is $H_L$.}
    \label{fig:comp_summary}
  \end{center}
\end{figure}

We consider an urban microcell deployment and focus on the canonical 3-cell 
topology shown in Figure~\ref{fig:sim_topo} with 3 mobile clients, one per each 
cell, that are being served at any time. 
We use a MATLAB implementation~\cite{bib:matlab} of the 3GPP Spatial Channel 
Model (SCM)~\cite{bib:3gppscm} for generating the time-varying $3\times 3$ 
channel matrices.
The simulation parameters are summarized in Table~\ref{tab:simpar}.
The coordination-based joint transmission procedure (CoMP-JT) is summarized in 
Figure~\ref{fig:comp_summary}. 
The coordination mechanism is abstracted out in Step 2 of the figure - the 
design of the coordination plane will be the focus of the next section, all that matters 
for the analysis in this section is the coordination latency $L$ which is treated as a variable. 
We make use of a practical and nearly optimal \tit{soft interference nulling} 
precoder~\cite{bib:sin} to precode user data before joint transmission.

\begin{figure}[htpb]
  \begin{center}
    \includegraphics[scale = 0.22]{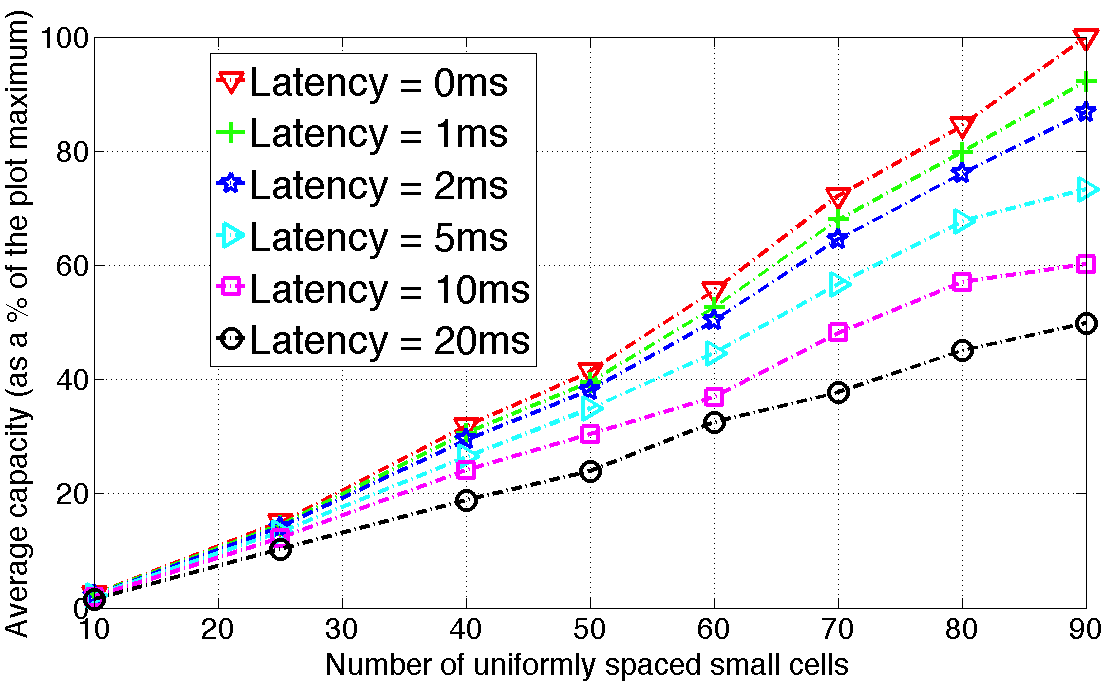}
    \scriptsize \caption{Average network capacity with CoMP vs (a proxy for) density. 
    Each point on the x-axis decreasingly corresponds to a microcell radius between 
    600 and 200m.}
    \label{fig:density}
  \end{center}
\end{figure}

Figure~\ref{fig:density} shows the average capacity scaling for different
latencies as a function of the number of uniformly spaced microcells required to
cover a circular region of 2km radius.  In other words, the horizontal axis
represents decreasing microcell radii which is a proxy for increasing density.
Note that these latencies represent the total delay between the time the
channels are estimated and the time they are used for precoding and
transmission, and are usually at least 1ms higher than the control plane latency
due to the feedback and other processing delays in between (so the Latency = 1ms
curve actually represents a zero-latency control plane and so on).

The small cell industry expects a linear increase in network capacity with the addition 
of each small cell base station.
With interference coordination techniques like CoMP, as the Latency = 0-2ms curves 
show us, the linear scaling is indeed possible with near-instantaneous coordination
(the scaling is actually more than linear because cell splitting provides significant diversity 
gains in addition to linear multiplexing gains).
However, the capacity scaling drops very quickly with density as the total latency exceeds 
5ms. 
Specifically, for the parameters used in this simulation, we find that the average network
capacity drops by as much as 40\% when the total latency is 10ms and 50\% when it is 20ms.

\begin{figure}[htpb]
  \begin{center}
    \includegraphics[scale = 0.27]{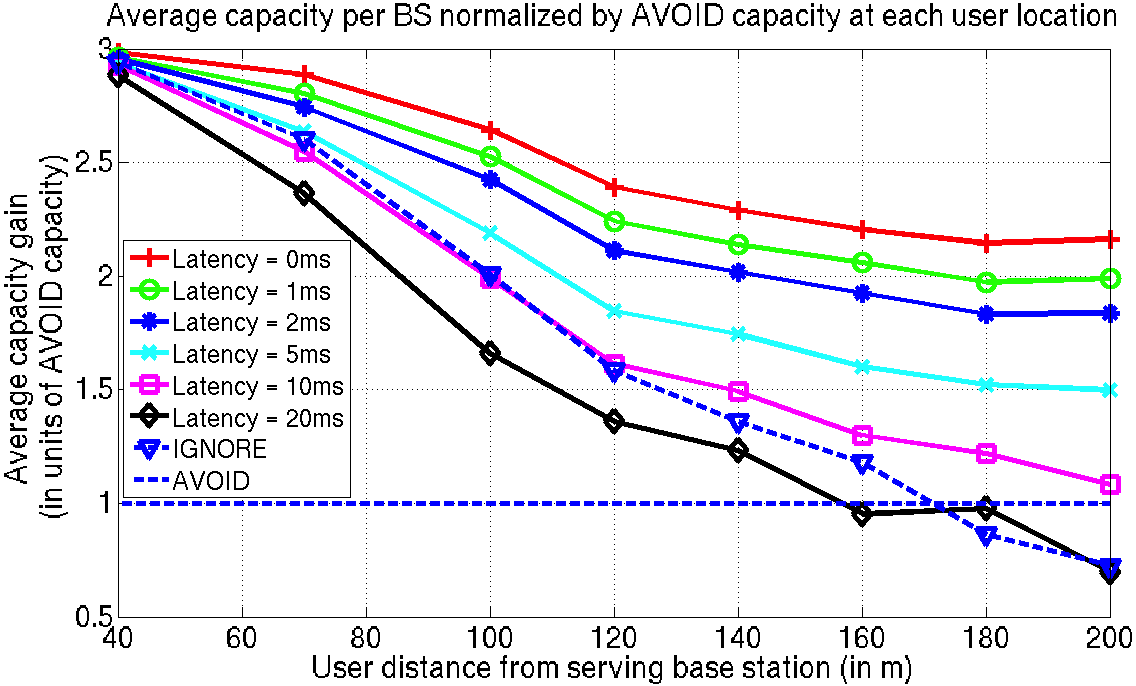}
    \scriptsize \caption{Average gains from coordination as a function of user distance
    from the serving base station. The capacity at each user location is normalized by
    the capacity that can be achieved by AVOID (i.e. time division between base stations).}
    \label{fig:test_distance}
  \end{center}
\end{figure}

We explore another question - does coordination at higher latencies provide any
gain at all over the simpler techniques of ignoring or avoiding interference?
If yes, how do these gains vary with user location (cell-core vs cell-edge)?  In
order to find an answer, we zoom into a specific microcell of radius 200m and
evaluate the average capacity achievable with CoMP-JT for different coordination
latencies as well as with \texttt{IGNORE} (i.e. treat interference as noise) and
\texttt{AVOID} (i.e. time division between the base stations).
Figure~\ref{fig:test_distance} plots the different average capacities per cell
normalized by the average \texttt{AVOID} capacity at each user location when the
users are symmetrically moved along the dotted lines in
Figure~\ref{fig:sim_topo}.	

Qualitatively, we observe that (1) the gains from coordination are significant only 
for edge users while \texttt{IGNORE} works almost as well for the core users. 
This makes sense because core users anyway see negligible interference 
(as compared to the direct signal strength), so coordinating with the neighboring 
base stations does not provide significant benefits.
(2) when the coordination latency is too high, the simpler technique of 
\texttt{AVOID} outperforms CoMP for edge-users.
(3) while coordination holds great potential for improving cell-edge throughput, 
it is also at the cell-edge where it is the most vulnerable to latency, indicated by
the largest percentage drop from the zero-latency capacity.
Conceptually, this happens because the interfering links are the strongest at the
edge, so any error due to staleness shows up as a large increase in the effective 
noise floor (the next subsection makes this notion clearer).

These observations and results collectively tell us that (1) interference
management techniques based on dynamic coordination, like CoMP-JT, can help us
realize the promise of linear network capacity increase with small cells, and
(2) in order to extract the maximum value out of such techniques, the
coordination latency inevitably needs to be on the order of a 2-3 milliseconds (the
lower the better).

\subsubsection*{Impact of coordination latency: An analytical view}

One might wonder if the drastic impact of coordination latency illustrated in
the previous subsection is an artifact of the specific choice of simulation
parameters. It is not.  In this subsection, we argue analytically that such
performance limits are in fact fundamental to dense urban networks. 

In order to analyze the impact of coordination latency, we need a model for how
the channels change over time.  We adopt a correlated block fading channel model
as described below.  Such a model, in addition to being analytically tractable,
captures the one essential piece needed to understand the impact of coordination
latency - the correlation between the current and the delayed CSI. 

Let $T_c(\rho)$ denote the $\rho\%$-coherence time of a single-tap channel (an OFDM
sub-channel is effectively single-tap). 
The correlated block fading model assumes that the channel stays constant over
a period of one $T_c(\rho)$, referred to as a \tit{slot}, and changes with correlation 
$\rho$ at the beginning of the next slot. 
Therefore,
\begin{equation}
 	h_{T_c(\rho)} = \rho \cdot h_0 + z_{T_c(\rho)}, |\rho| \leq 1
	\label{eq:time_model}
 \end{equation}
where $z_{T_c(\rho)}$ is independent of $h_0$ and $\Ex[z_{T_c(\rho)}]$ $=$ 0 
given that the complex baseband channels are zero-mean at all times. 
Assuming the channel is wide sense stationary, if $\sigma_h^2$ $\triangleq$ $\Ex[|h_0|^2]$, 
we must have $\Ex\big[|h_{T_c(\rho)}|^2\big]$ $=$ $\Ex\big[|h_0|^2\big]$, and 
therefore $\Ex\big[|z_{T_c(\rho)}|^2\big]$ = $(1 - \rho^2)\sigma_h^2$. 
Note that $\rho$ is a parameter of this block fading model and controls the 
granularity of discretizing the continuous-time channel - the closer $\rho$ 
is chosen to 1, the smaller will be $T_c(\rho)$ for a given channel and the finer 
will be the modeling of the channel variations. 

Using this model, if $L$ represents the latency of coordination between any pair 
of base stations (in order to keep the expressions simple, we will consider $L$ 
in multiples of $T_c(\rho)$), then the channel $h_L$ at $t = L$ is related to the 
channel $h_0$ shared at $t = 0$ as 
\begin{equation}
	\label{eq:latency_model}
	h_L = \rho^{L/T_c(\rho)}h_0 + z_{L}
\end{equation}
where $\Ex[|z_L|^2] = \big(1 - \rho^{2L/T_c(\rho)}\big)\sigma_h^2$. 

The term $\rho_L \triangleq \rho^{L/T_c(\rho)}$ represents the channel correlation 
across a time interval $\Delta t = L$ and measures how correlated the \textit{stale} 
CSI is to the actual channel at $t=L$. 
Let us define $S_L\triangleq (1-\rho_L)$ as the \textit{staleness} at $t=L$ of the CSI 
that was shared at $t=0$. 
As an example, if the current channel is 90\% correlated with the channel at some previous 
time, then by the above definition the previous CSI will be said to be 10\% stale currently.
The variance of the error due to staleness can thus be expressed as $(1 - (1-S_L)^2)\sigma_h^2$.

Now we are all set to argue why the impact of a given coordination 
latency $L$ is much more pronounced in dense urban networks.
In urban areas, the coherence times are typically smaller due to the inherent 
dynamism in the environment itself, which means that for a given $L$, the staleness 
$S_L = 1-\rho^{L/T_c(\rho)}$ ($|\rho|\leq 1$) is higher. 
In addition, if the network is dense, the links are typically stronger (since the clients 
are closer to the base stations), and hence  $\sigma_h^2$ is higher.
Therefore, in dense and urban networks, the problem gets compounded and the 
error $(1 - (1-S_L)^2)\sigma_h^2$ due to a given latency $L$ is much higher.
As the network becomes denser, the higher errors in CSI lead to higher errors in the precoding
vectors which in turn lead to the huge capacity drops that we saw earlier in Figures~\ref{fig:density}
and~\ref{fig:test_distance}.

We would also like to clarify here that, for example, a 10\% staleness (or equivalently,
90\% correlation) is not the same as a 10\% error in channel coefficients. 
As we explained above, a 10\% staleness actually results in a $\big(1 - (1-0.1)^2\big)$
or 19\% error relative to the channel coefficients.

\section{Design}
\label{sec:design}

\name is a control plane design for heterogenous and dense cellular networks consisting
of macro and small cells capable of providing moderate bandwidths (on the order
of an Mbps) at very low latencies (on the order of a millisecond). 
Its main component is a novel design for low-latency physical backhaul connectivity 
to the macro base station for control signaling that enables the use of
CoMP and other latency-sensitive network management mechanisms. 

\name's key architectural insight is to \emph{physically} decouple the LTE data
and control planes in the backhaul. As we saw before, the LTE control plane
requires extremely low latencies on the order of 2-3 ms, but is not bandwidth
intensive. The LTE data plane on the other hand is less latency sensitive
(latencies on the order of 20-30ms are fine for most data traffic and even VoLTE
voice calls), but is quite bandwidth sensitive (requires several hundreds of
Mbps backhaul connectivity). By physically coupling the data and control planes
(i.e. using the same physical backhaul connectivity) and layering both control
and data signaling on top, we end up with a situation where the small cell
backhaul has to provide \emph{both} very high bandwidth and extremely low
latency. Such a stringent requirement limits the backhaul options available to
operators, they are forced to use direct fiber links since thats the only
technology that can provide the combination of high bandwidth, low latency and
reliability that LTE networks need. Such fiber deployments however are
prohibitively expensive (imagine deploying fiber in an urban environment such as
Manhattan), and have prevented small cell deployments from happening at scale.

\name physically decouples the LTE data and control planes. The key insight is
to use the existing LTE macro network itself as the physical layer technology
for control plane backhaul. In other words, \name's control plane signaling
happens on LTE frequencies itself between the macrocell and the small cells. By
doing so, small cells gain a direct connection to the macro-cell control plane
coordinator and an extremely low latency link (on the order of 2-3ms). Operators
are then free to choose data plane backhaul that is high bandwidth but doesnt
have to satisfy the stringent latency requirements. Such backhaul options are
more easily available in the form of cable networks, ADSL and even public fiber
networks switched over another operator's wireline networks.

\name does not require additional frequency bands as it creates a control 
plane within the LTE spectrum itself. However the challenge is that we may now
be using extremely scarce and expensive LTE spectrum for control signaling. The
key contribution of this paper is a design that enables a very efficient control
plane design using existing LTE spectrum. 

\name's control plane logical architecture is similar to current LTE deployments
and hence it can be easily deployed. As before each small cell connects to a
logically centralized coordinator which is deployed at the macrocell within
whose coverage area the small cells are located. The small cells use \name
to send and receive control signals from the coordinator. The coordinator is
reponsible for making all radio network management decisions.

\subsection{\name Backhaul Design}
\label{subsec:backhaul}
The key idea in designing \name's control plane backhaul is to use the existing LTE 
spectrum itself for control signaling with the macro. 
In other words, each small cell sends and receives control messages (such as channel 
state measurements, decisions on which small cells should coordinate using CoMP and 
how etc.) to and from the coordinating macro on the same LTE spectrum that is used for 
data transmission.
To implement this, each small cell is equipped with a UE (phone) radio, which it uses to 
connect with the macro base station just like any other phone. 
It uses this connection to send and receive control messages.

Such a design has several attractive properties. 
First it is extremely simple to deploy, since a macro network is already architected to talk 
to phones, adding control links to each small cell is trivial. 
Second, it works quite well in urban non line-of-sight environments because it uses LTE frequencies 
and physical layer protocols which were designed to work well in such environments.
Further it requires no additional spectrum unlike other microwave backhaul based solutions. 

A natural question however is whether we are using valuable LTE spectrum for control signaling. 
To see why this might happen, we have to look at how an LTE small cell operates. 
We focus on LTE FDD (frequency division duplexing)~\cite{bib:3gppphy} small cells in this paper, 
but the same arguments apply for TDD LTE (Time Division Duplexing). 
FDD LTE has two frequencies, one for the uplink which is used by the UEs to transmit to the eNB 
(both macro and small cells), and the other is the downlink which is used by the eNB to transmit to the UEs. 
However with \name's design, a small cell eNB also acts as a phone which connects to the macro eNB. 
This implies that the small cell may have to transmit on the uplink frequency to the macro eNB while 
it is receiving data from the UEs that are connected to it on the same uplink frequency, and the reverse 
on the downlink frequency. 

A natural design choice is to separate the control transmissions/receptions from
the small cell eNB to the macro in time with the data transmissions/receptions
from the UEs connected to the small cell. In other words turn the small cell
into a half duplex LTE mesh node to coordinate transmit and receive. However
this can be extremely wasteful. For example to implement CoMP in an urban
deployment, small cells have to exchange channel state information with the
macro eNB at a rate of 5Mbps or so (assuming walking mobility and the need to
send updates at least every 2ms). While the small cell is sending this information to
the macro eNB, it cannot receive from its own UEs and throughput is
significantly reduced. In fact, we find that sending such control information means 
that the small cell cannot listen to its own UEs for 25\% of the time. 
Further such scheduling in the small cell is extremely complex
to build because in effect the small cell has to schedule its own transmissions
and receptions around when the macro eNB schedules the small cell to
transmit/receive control messages to/from it. 

\begin{figure}[htpb]
  \begin{center}
    \includegraphics[scale = 0.23]{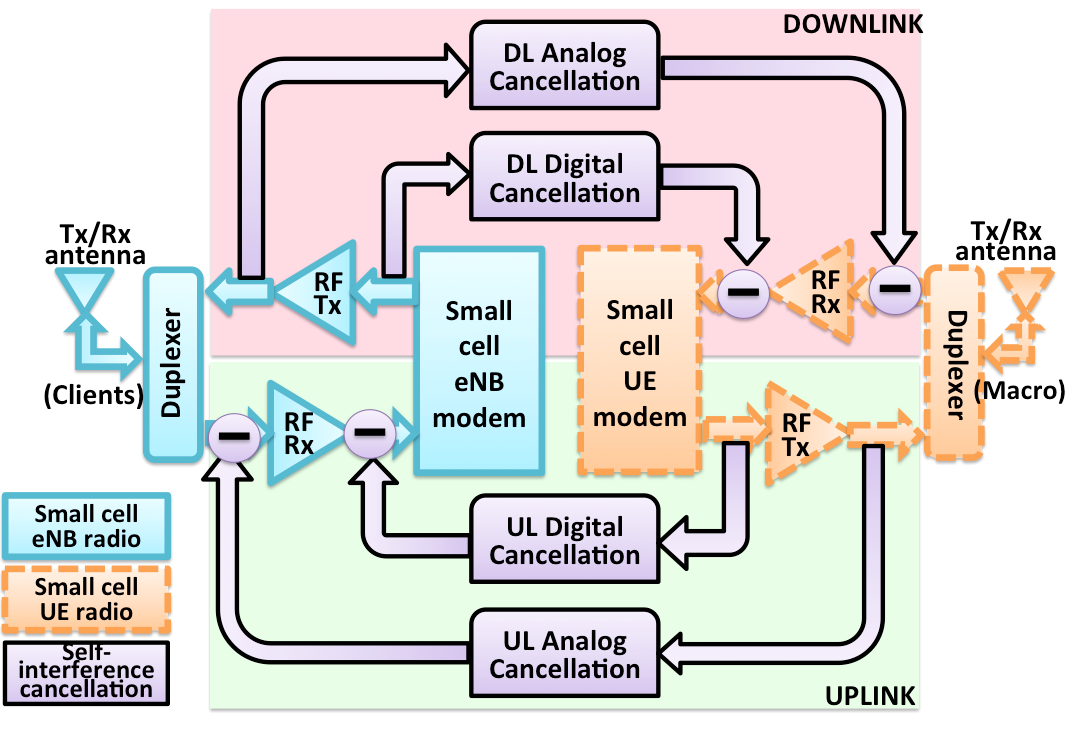}
    \caption{Block Diagram: A small cell equipped with \name. To the 
    existing eNB radio, \name adds an UE radio and self-interference cancellation units
    for uplink and downlink. As a result, a \name small cell can act as an eNB on one 
    side (send/receive data to/from clients) and UE on the other (send/receive control
    to/from a macro coordinator).}
    \label{fig:arch}
  \end{center}
\end{figure}

Our key innovation is to leverage recent work on full duplex radios to turn each
small cell into a dual full-duplex radio that can act simultaneously as an eNB
(to its clients) and UE (to a macro eNB), see Figure~\ref{fig:arch}. 
In other words, a small cell equipped with \name would be able to send control
information to a macro eNB while listening to its own clients on the uplink, 
and at the same time would also be able to receive control signals from the macro eNB 
while serving its own UEs on the downlink.
Thus, the \name control plane can effectively coexist with the small cell user plane 
without hurting its capacity. 
In order to enable this coexistence, \name must cancel the self-interference caused 
by the transmission of the small cell eNB at its UE receiver (i.e., on the downlink), 
and similarly the transmission of its UE at its eNB receiver (i.e, on the uplink), as 
shown in Figure~\ref{fig:arch}.

\subsubsection*{How much self-interference cancellation is needed?}
\label{subsubsec:cancellation}
Each of the uplink and downlink self-interference canceling units must be able to 
reduce the maximum transmit power to the noise floor at the corresponding receiver 
over the entire bandwidth of operation (any residual interference below the noise floor 
does not affect performance). 
Consequently, the total self-interference cancellation required for each Tx-Rx pair i.e.,
uplink Tx-Rx and downlink Tx-Rx, is the difference between the \tit{maximum Tx power} 
and the \tit{Rx noise floor} (expressed in dBm). 

\tbf{Rx noise floor:} The receiver noise floor depends on the bandwidth B and the noise figure 
NF of the receiver, and can be estimated as -174dBm/Hz (thermal noise at 290K) + 10log$_{10}$(B) 
+ NF (in dB), see Table~\ref{tab:noise_floor}.
\begin{table}[h]
	\begin{center}
		\caption{Typical receiver noise floors}
		\begin{tabular}{|c|c|c|c|c|}
		\hline
	   					& 			& \multicolumn{3}{c|}{Receiver noise floor (dBm)}	\\ \cline{3-5}
		Receiver			& NF			& 5MHz		& 10MHz			& 20MHz		\\ \hline
		\tbf{Small cell eNB}	& 5dB		& $-$102		& $-$99			& $-$96		\\ \hline	
		\tbf{Small cell UE}	& 9dB$^*$	& $-98$		& $-$95			& $-$92		\\ \hline
		\end{tabular}
		\label{tab:noise_floor}
	\end{center}
	{\small $^*$assuming it to be the same as for a conventional UE}
\end{table}

\begin{figure}[htpb]
  \begin{center}
    \includegraphics[scale = 0.22]{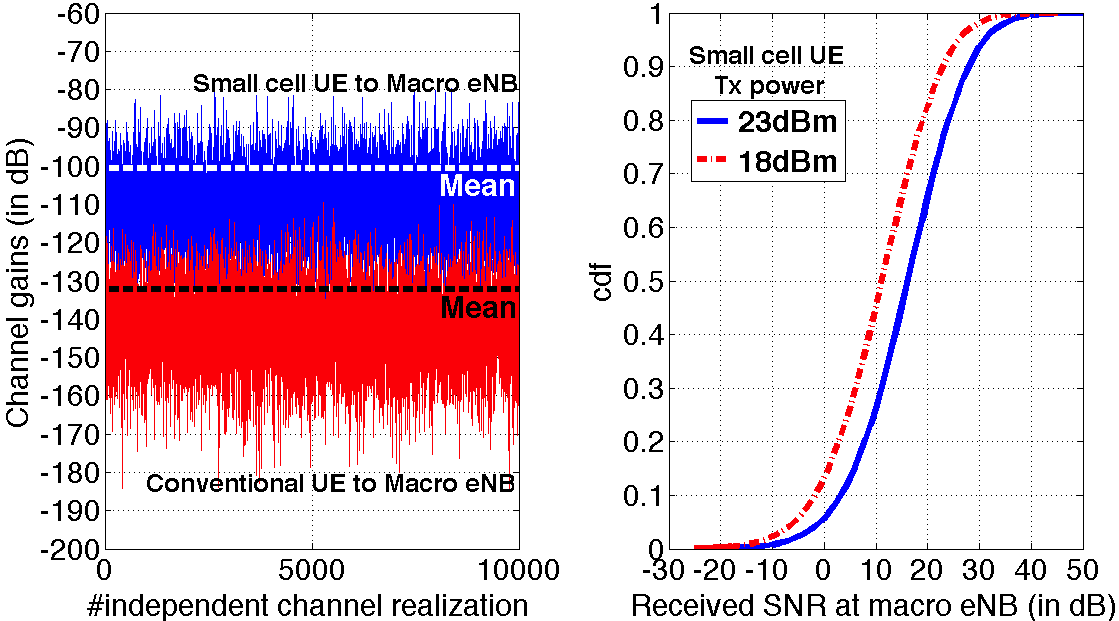}
    \caption{(Left) An illustration of channel gains seen by a small cell UE (12.5m high)
    and a conventional UE (1.5m high) to a macro eNB (32m high) located 1km away. 
    (Right) Cdf of Rx SNR at the macro eNB for two different small cell UE Tx powers
    (23dBm and 18dBm).}
    \label{fig:channelUEcdf}
  \end{center}
\end{figure}

\tbf{Max Tx power}: 
\tit{First}, on the uplink, we observe that a small cell UE can be expected to experience much 
lower path losses to a macro eNB than a conventional UE since it would typically be located 
10m or higher above the ground, and can therefore afford to transmit at a lower power.
In fact, using the 3GPP Spatial Channel Model (SCM)~\cite{bib:3gppscm}, we find that the 
channels seen by an urban microcell UE (12.5m high) located at the edge of a 1km-radius 
macrocell (32m high) can be expected to be as much as 30dB stronger on average than by 
a conventional UE (1.5m high) at the same distance, see Figure~\ref{fig:channelUEcdf} (Left).
We also find that at 23dBm, the maximum specified Tx power for UEs, the Rx SNR at the 
macro eNB is nearly 50\% likely to be in excess of the maximum useful SNR ($\sim$18dB, 
corresponding to the highest MCS for uplink i.e., 64QAM and 0.85 code), thereby leading to a 
wastage of power, see Figure~\ref{fig:channelUEcdf} (Right). 
Instead, if the small cell UE lowers its Tx power by 5dB, it can reduce the likelihood of power 
wastage by 25\% and yet stay connected for more than 95\% of the time (note that these are 
the worst case estimates - the small cell is located farthest from the macrocell).
Thus a practical estimate for the maximum Tx power of a small cell UE is 18dBm.

\tit{Second}, on the downlink, we believe the expected transmit power used will be around 30dBm. 
The reason is that typical small cell deployments are for covering hotspots such as bus-stops, and
hence the coverage area required is small. 
Further small cells are expected to be deployed on light poles, walls etc, and hence cannot afford 
to be big and heavy. 
Using transmit powers higher than 30dBm results in big and hot power amplifiers that require large
boxes to cool them and make it infeasible to build a ``small" cell. 
Therefore most actual small cell deployments will have compact base stations that have a transmit 
power around 30dBm (1 watt). 
\squeezeup

Table~\ref{tab:cancellation} summarizes the self-interference cancellation requirements
for \name. It is important to mention here that although wider bandwidth systems require
lesser cancellation owing to their higher noise floor, in practice it is more difficult to match 
the cancellation blocks over wider bandwidths and hence, self-interference cancellation 
is practically much harder in wider bandwidth systems.
\begin{table}[h]
	\begin{center}
		\caption{\name: Cancellation requirements}
		\begin{tabular}{|c|c|c|c|c|}
		\hline
	   				& Max				& \multicolumn{3}{c|}{Cancellation required (in dB)}	\\ \cline{3-5}
		Band		& P$_{\text{transmit}}$	& 5MHz		& 10MHz			& 20MHz		\\ \hline
		\tbf{Uplink}	& 18dBm				& 120		& 117			& 114		\\ \hline	
		\tbf{Downlink}	& 30dBm$^*$			& 128		& 125			& 122		\\ \hline
		\end{tabular}
		\label{tab:cancellation}
	\end{center}
	{\small $^*$assuming the small cell is a microcell}	
  \end{table}

\subsubsection*{How does \name achieve this cancellation?}
\name leverages recent work on full-duplex radios~\cite{bib:full-duplex} and employs 
a combination of analog and digital cancellation blocks to provide the required 
self-interference cancellation on both uplink and downlink frequencies. 
In a nutshell, the self-interference in each 
Tx-Rx pair consists of 
(1) about 65-70dB of transmitter (broadband) noise generated by the components of the 
analog RF Tx front end, which must be necessarily canceled in the analog domain (by 
taking a copy from where it is generated), and 
(2) about 50-55dB of the remaining Tx signal which consists of nearly 20dB of residual 
non-linearities (after analog cancellation) and 30-35 dB of linear components, and can 
be canceled in the digital domain. 
We refer the readers to~\cite{bib:full-duplex} for the details of the algorithms used to 
design the analog and the digital cancellation blocks that can meet the above requirements.
In Section~\ref{sec:eval}, we build a prototype and demonstrate that \name can indeed 
provide the required self-interference cancellation with only a marginal (1.7dB) increase 
in noise floor.

\subsection{Increasing efficiency}
\label{subsec:efficiency}
One of the major concerns with \name is whether it expends a lot of scarce LTE spectrum
for coordination. 
We have described how \name can leverage full duplex radios to ensure that no capacity 
is lost at the small cell.
However, \name does consume some macrocell resources both on the uplink and the
downlink, and addressing this overhead is a key aspect in the design of \name. 
In the following discussion, we show how we can exploit the properties of \name's physical 
backhaul medium itself, namely the LTE physical layer, to increase the efficiency of \name.

\subsubsection*{How can we optimize the overhead on macro resources?}
\tit{First}, we note that the channels a macro eNB would see to the small cell UEs over 
\name can be expected to be \tit{much more static} and of \tit{much better quality} than 
to mobile UEs located at comparable distances.
This is because the small cells are typically installed 10m or higher above the ground 
and hence their path losses are significantly lower~\cite{bib:3gppscm}, see 
Figure~\ref{fig:channelUEcdf} (Left).
In addition to making a case for introducing higher modulation and coding schemes for
this special category of small cell UEs to exploit the naturally good channel quality, this
offers the advantage of \tit{requiring much lesser time-frequency resources for communicating 
a given amount of control information}.
For e.g., using 3GPP SCM~\cite{bib:3gppscm}, we find that from the perspective of a 
macrocell, providing 1 Mbps of average uplink throughput to a small cell UE located at the 
edge (1km away) is equivalent to denying only 183.4 Kbps of uplink throughput to a UE located 
just 500m away and just 39.0 Kbps to a UE located at the cell-edge.
Thus, using the wireless physical layer as the backhaul medium works to \name's advantage
by \tit{lowering the effective overhead} it incurs on macrocell capacity.

\tit{Second}, we observe that depending on the control information, which mostly relates to 
some form of channel state information (recall that \name is intended for communicating control
information that changes on the order of 1ms), there exists a scope for exploiting redundancies 
in the control information itself to \tit{optimize the overhead}.
For example, when a small cell has to forward the CSI of its clients over an LTE resource block 
(consisting of 12 sub-channels) to the macro coordinator, it can exploit the strong correlation 
between the narrow-band LTE sub-channels ($\sim$15KHz) to compress the information.
Similarly, even though the channels decorrelate much faster over time in dense urban environments, 
we find that non-trivial compression can also be achieved across time by scheduling 
clients over consecutive sub-frames on the same sub-channels.
 
In essence, if $\vec{h}^T$ = $[h_0, h_2, \cdots, h_{N-1}]$ represents the channel coefficients 
seen by a client to a small cell eNB across $N$ consecutive sub-channels in a given sub-frame 
(or $N$ consecutive sub-frames over a given sub-channel), then the small cells can transform
$\vec{h}$ into another domain using an appropriate sparsity transform $T$ such that the resulting 
sequence $\widetilde{\vec{h}} = T\vec{h}$ is sparse. 
As a result, $\widetilde{\vec{h}}$ can be represented using much fewer bits than $\vec{h}$ for a 
given level of distortion. 
Note that these compressions can also be performed for \name by the clients themselves before
feeding back their CSI to the small cells, but that might require firmware changes to the current 
LTE UEs.
Since we want to keep the design of \name independent of UEs, we let the small cells optimize 
the representation of the control information before sending over \name.
\begin{figure}[htpb]
  \begin{center}
    \includegraphics[scale = 0.22]{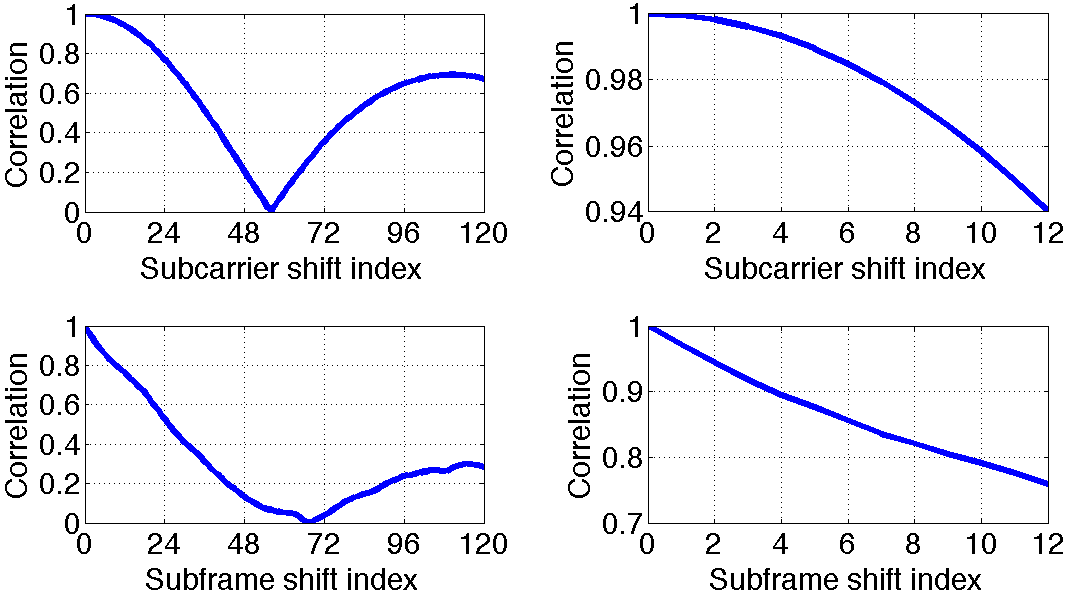}
    \caption{Estimates of correlation across adjacent sub-channels (top) and sub-frames (bottom)
    for a randomly placed UE, obtained by averaging over a particular realization of the channel
    over 512 sub-channels and 500 sub-frames. The plots on the right are zoomed-in versions of
    the plots on the left and show a strong correlation over a block of 12 time/frequency slots.}
    \label{fig:correlation}
  \end{center}
\end{figure}

Although we do not deal with optimal sparsity transforms in this paper, we show in Section~\ref{sec:eval} 
that the following simple and practical encoding scheme can reduce the overhead of \name to a 
negligible fraction of the macrocells capacity on the uplink.
If $q$ bits of magnitude and $q$ bits of phase are required to reliably represent a channel coefficient, 
the Channel Quality Indicator CQI and Channel Phase Indicator CPI can belong to one of $2^q$ 
quantization levels.
In order to encode and send each coefficient along consecutive sub-channels or sub-frames, 
send the first coefficient of the series $h_0$ using $q+q$ bits, and then send the CQI of subsequent 
coefficient $h_n$ as (1) bit \texttt{0}, if CQI$_n$ = CQI$_{n-1}$, (2) bits \texttt{10}, if CQI$_n$ $-$ 
CQI$_{n-1}$ = -1, (3) bits \texttt{11}, if CQI$_n$ $-$ CQI$_{n-1}$ = 1, or (4) $q$ bits otherwise
(and similarly for CPI) .

This scheme essentially encodes the \tit{increment} using Huffman coding if the increment is within
one quantization level, otherwise it \tit{refreshes} it i.e., sends the full $2q$ bits for the coefficient.
In Section~\ref{sec:eval}, we show that this scheme works very well in practice because for typical 
LTE OFDM channels, the increments over one sub-channel or sub-frame is less than one quantization 
level with a very high probability (with an increment of 0 being more likely than an increment of $\pm1$).

In addition, we note that there typically exists significant intra-site correlation between 
the channels of a user to the different antennas of a small cell as well as inter-site correlation 
between the channels seen to neighboring cells~\cite{bib:3gppscm}, which can be exploited 
similarly for lowering the overhead further, if needed. 

Also, we note that the control information on the downlink will usually be a few bits per interface
per sub-frame, usually representing a decision index or a network state indicator, for e.g. PMI
for interference management.
Combined with the facts that a macro eNB typically has a much larger transmit SNR and sees 
very strong channels to the small cell UEs, we believe that \name will have a negligible impact 
on the overall downlink capacity of the macrocell.

\subsection{Illustration: Interference management with \name}
Now that we have described the design of \name, we take up a typical example of
a dense cellular network consisting of a macrocell and several smalls
and illustrate how \name can be used for interference coordination in small cells.
\begin{figure}[htpb]
  \begin{center}
    \includegraphics[scale = 0.23]{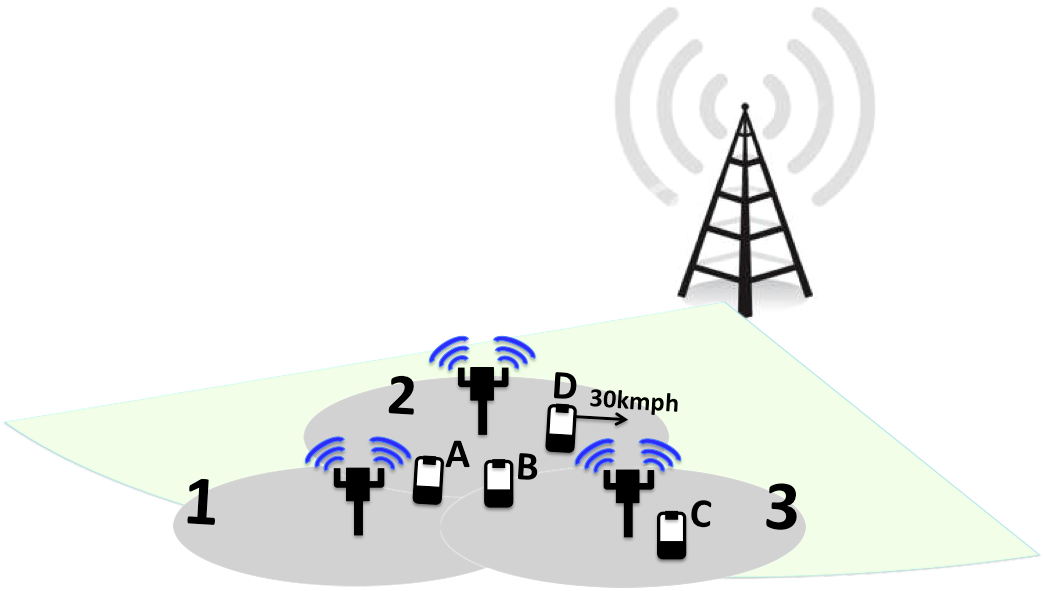}
    \caption{Illustration of an urban hetnet. A 3-microcell cluster (equipped with \name) is 
    deployed in a sector of a tri-sector macrocell.}   
    \label{fig:hetnet}
  \end{center}
\end{figure}
\squeezeup

Figure~\ref{fig:hetnet} illustrates a variety of interference scenarios in an urban microcell cluster 
deployed within a sector of a 3-sector macrocell.
Clients A and B are edge users in their respective microcells and see strong interference from 
their neighboring micro eNBs.
Client C lies in the core of microcell 3 but might be seeing strong interference from the macro eNB.
Client D is a high mobility client who happens to be within the coverage area of microcell 2 at this
instant.

A typical network interference management scheme might choose to
\begin{Enumerate}
	\item let client D be served by the macro eNB even though it lies within the microcell cluster 
	since it is not going to stay in the cluster for long anyway.
	\item let client C be served by microcell 3, and \texttt{AVOID} or reduce interference from the 
	macro eNB (for e.g., eICIC) if its interference is strong, otherwise \texttt{IGNORE} interference.
	\item \texttt{EXPLOIT} interference for client A using CoMP joint transmission by microcells 1 and 2.
	\item \texttt{EXPLOIT} interference for client B using CoMP joint transmission by all three microcells.
\end{Enumerate}

\begin{figure}[htpb]
  \begin{center}
    \includegraphics[scale = 0.22]{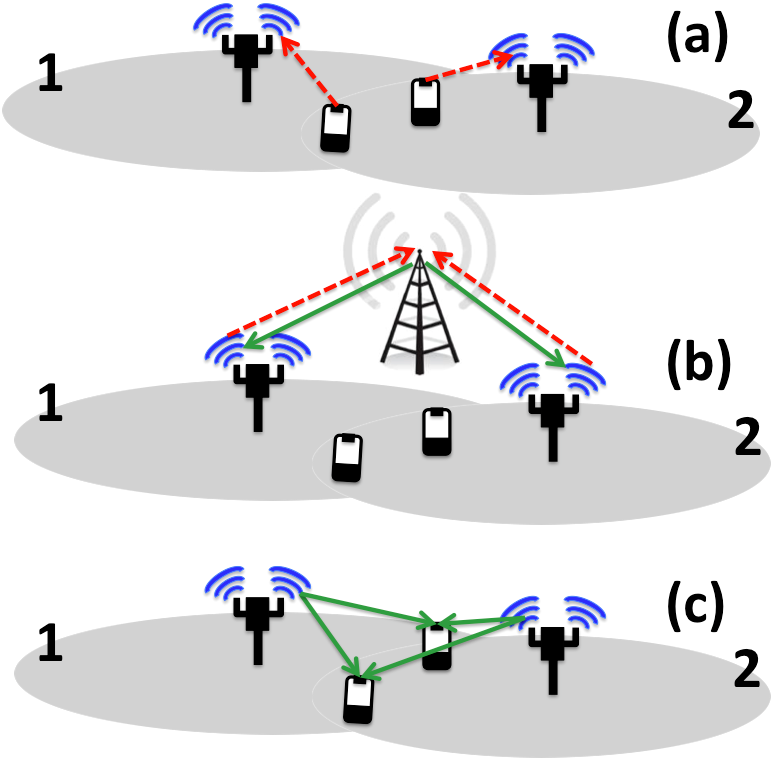}
    \caption{Illustration of coordination over \name.}
    \label{fig:coordination}
  \end{center}
\end{figure}

While clients C and D can be served with little or no control signaling between the different eNBs,
the microcells need to coordinate on the order of a millisecond to serve clients A and B using CoMP.
If the microcells are equipped with \name, this low-latency coordination is simple to execute.
Figure~\ref{fig:coordination} summarizes the coordination mechanism for a $2\times 2$ scenario.
(a) The clients report to their serving micro eNBs on the uplink their respective channels to the two 
micro eNBs. 
(b) The micro UEs process and forward the compressed CSI via \name on the uplink. 
The macro eNB replies back with coordination parameters (PMI, powers, schedules etc.) on the downlink. 
All this time while the micro UE is communicating with the macro eNB, \name's self-interference 
cancellation blocks ensure that the micro eNB can keep communicating with its other clients as usual 
both on the uplink and the downlink.
(c) The micro eNBs precode and jointly transmit to the clients on the downlink.
\squeezeup

\subsection{\name: Latency Guarantees}
Since \name uses LTE data channels to carry control traffic, it inherits the user-plane latency
that LTE itself guarantees. 
However, there are a few significant differences that promise to make \name's latency guarantees 
even better. 
\tit{First}, since \name carries critical control information, the macro eNB can prioritize this traffic
including pre-allocating resource blocks, which would mean lesser processing delays and lesser
waiting times.
Consequently, \name can provide a best-case one-way latency of 1 TTI (Transmission Time Interval),
which is 1ms in current LTE standards, and an average one-way latency of 1.5 TTIs that includes an
average waiting time of 0.5 TTI until the start of the next sub-frame.
\tit{Second}, as the wireless channel that \name operates over is significantly stronger (owing to 
lower path losses at the heights at which small cells are typically installed), \name traffic can be
expected to suffer lesser re-transmissions and can consequently guarantee a better average latency.
\tit{Third}, our simulations have shown the typical control packets to be lesser than 0.5ms in duration, 
so the latency of \name is largely limited by the TTI defined in current standards and will automatically 
improve as the standards evolve to support shorter transmissions.

\begin{figure}[htpb]
  \begin{center}
    \includegraphics[scale = 0.22]{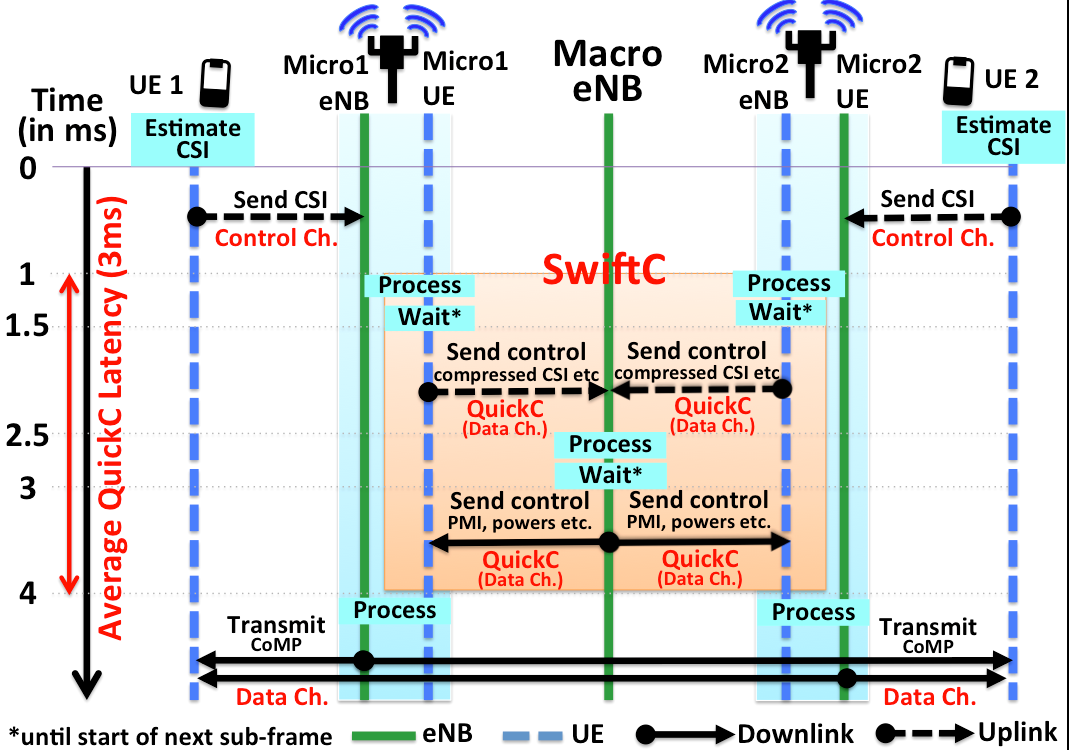}
    \caption{Example of a coordination timeline (cf. Figure~\ref{fig:coordination}).}
    \label{fig:timeline}
  \end{center}
\end{figure}

Since one use of the control plane involves one use each of \name's uplink and downlink, \name 
can provide a best-case latency of 2 TTIs (= 2ms in current standards) and an average two-way
latency of 3 TTIs. 
The total coordination latency however includes the sum of the feedback and processing delays, 
and can be expected to be 1-2ms higher.
Figure~\ref{fig:timeline} shows a typical coordination timeline for the example illustrated in
Figure~\ref{fig:coordination}.
\squeezeup

\section{Evaluation}
\label{sec:eval}
In this section, we
\begin{Enumerate}
	\item design a prototype of \name's backhaul and demonstrate that it can provide 
	the highest self-interference cancellation required at the widest bandwidth supported 
	by LTE (namely 122dB at 20MHz).
	\item evaluate the bandwidth requirements of the \name plane and show that it 
	translates to a negligible overhead in terms of the overall macro capacity.
	\item simulate the performance of interference coordination techniques such a CoMP 
	in a dense LTE macrocell (containing several microcells) and show that the overall
	network capacity using \name can scale linearly with density ($N\times$ increase with
	$N$ small cells), something that the current X2 over IP control planes fall way short of providing.
\end{Enumerate}

\subsection{Implementation of \name}
We design and build a prototype of \name to evaluate its novel physical backhaul
design for the LTE control plane. The goal is to show that it is possible to
implement a full duplex LTE small cell that connects to the macro cell as a UE
for sending and receiving conrol plane traffic. The primary determinant of this
is whether \name can implement the self-interference cancellation needed to
build the full duplex capability with the small cell eNB and the UE.

\begin{figure}[htpb]
  \begin{center}
    \includegraphics[scale = 0.23]{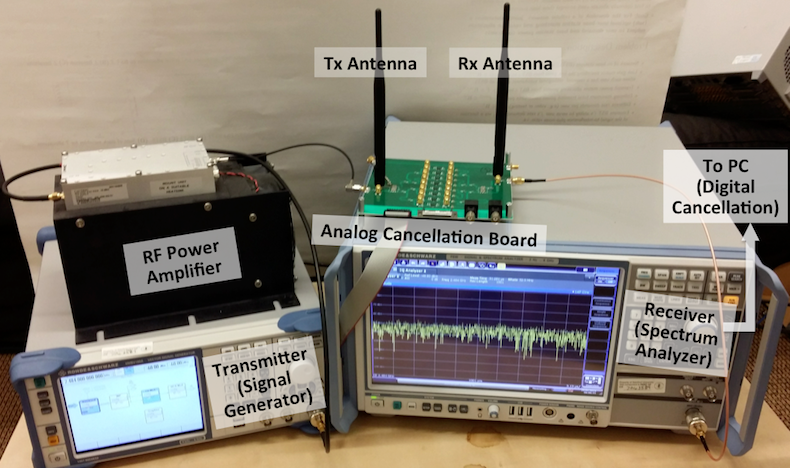}
    \caption{Experimental setup of a \name Tx-Rx pair.}
    \label{fig:setup}
  \end{center}
\end{figure}

To experimentally verify this, we build a prototype with the following components,
see Figure~\ref{fig:setup}:
\begin{Itemize}
	\item a \tit{transmitter} (implemented using a vector signal generator) that transmits 
	standard OFDM signals at 2.4GHz over 20MHz, the widest bandwidth that LTE supports.
	\item an \tit{RF power amplifier} that boosts the transmit power to 30dBm,
	the maximum that a \name-equipped small cell radio is expected to transmit at.
	\item a \tit{Tx antenna} and an \tit{Rx antenna} that act as the \name Tx-Rx pair,
	and correspond respectively to the antennas on the small UE radio and the 
	small cell eNB radio on the uplink and vice-versa on the downlink.
 	\item an \tit{analog cancellation board} which implements fixed delay lines and 
	programmable attenuators that can execute the analog cancellation 
	algorithm proposed in~\cite{bib:full-duplex}.
	\item a \tit{receiver} (implemented by a spectrum analyzer) that converts the analog 
	received signal into digital I-Q samples. 
	The widely-used software radios such as USRPs and WARPs typically have very poor 
	noise floors ($\sim-85$dBm over 20MHz) and do not let us demonstrate the functioning 
	of \name for typical UE and eNB radios whose noise floors are at least 5-7dB better, 
	see Table~\ref{tab:noise_floor}.
	Therefore we use a commercial radio test equipment (-90dBm noise floor at 20MHz) 
	for implementing our receiver.
	\item a \tit{digital cancellation block} implemented in software (MATLAB) using the 
	algorithm proposed in~\cite{bib:full-duplex} that acts on I-Q samples from the spectrum 
	analyzer.
\end{Itemize}  

Our goal here is to show that \name can provide the required self-interference cancellation
required on both uplink and downlink over the widest bandwidth (20MHz) that LTE supports.
Note that self-interference cancellation is more difficult when the bandwidth is larger since 
matching the response of the cancellation blocks to wider bandwidths is practically much harder.

\begin{figure}[htpb]
  \begin{center}
    \includegraphics[scale = 0.5]{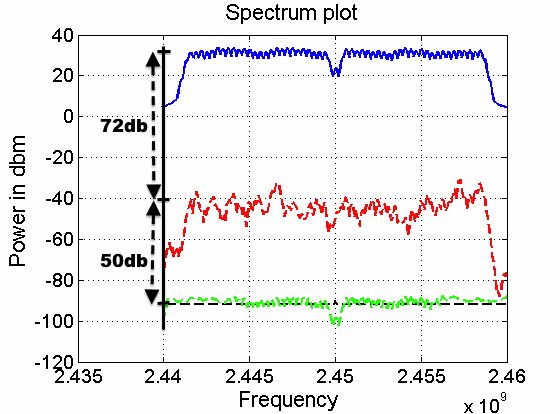}
    \caption{Self-interference cancellation provided by \name prototype. The 32dBm transmit
    signal over 20MHz is reduced in steps by 72dB (analog cancellation) and 50dB (digital 
    cancellation) to the noise floor at -90dBm.}
    \label{fig:expcancellation}
  \end{center}
\end{figure}

Figure~\ref{fig:expcancellation} shows the cancellation performance of our prototype.
The transmit signal is a standard LTE-like OFDM signal at 2.4GHz and 32dBm (35dBm at power 
amplifier output followed by a 3dB cable loss). 
The analog cancellation board reduces the signal to -40dBm (72dBm cancellation) while
the digital cancellation board clears up another 50dB of self-interference to reduce it to the 
noise floor. 
The result is 122dB of total cancellation, the maximum that \name needs to provide at 20MHz,  
see Table~\ref{tab:cancellation}. 
The effective increase in noise floor, as measured from the samples after digital cancellation,
is 1.7dB. 

Note that the cancellation requirement on the uplink is smaller (only 114dB at
20MHz) and can be achieved by this same setup, although we omit any experimental
results for the uplink since our radio test equipment cannot emulate the noise
floors of commercial eNB radios.  In the rest of this section, we treat the
self-interference cancellation blocks as a blackbox that raises by the noise
floor at the small cell eNB and the small cell UE radios by 1.7dB. Next we turn
to simulations to evaluate the performance of the \name control plane in
practical LTE small cell deployments. The simulations use the same deployment
conditions and channel models that are used by the LTE standards body to
evaluate new proposals for standardization, and hence are considered to be quite
representative. Further, its infeasible except for the largest operators to
actually be able to deploy a full scale system implementation of \name and
evaluate it under realistic conditions. Hence we turn to detailed simulations.

\subsection{\name benchmarks}
In this subsection, we benchmark \name on the following two aspects.
\begin{Enumerate}
	\item How much control traffic does \name need to support, assuming that resources 
	would need to be coordinated on per 1ms sub-frame basis and that CSI is going to 
	be the dominant form of control information? (Section~\ref{subsec:traffic})
	\item What overhead does this traffic incur on the uplink capacity of the macrocell? 
	(Section~\ref{subsec:impact_macro})
\end{Enumerate}

\subsubsection{Bandwidth requirements}
\label{subsec:traffic}
In order to estimate the bandwidth requirements of \name, we analyze from the perspective 
of a single small cell. 
A small cell serves its clients by allocating each of them one or more resource blocks (RBs),
each consisting of 12 sub-channels, for one or more sub-frames.

Let us look at how many bits would be needed to represent the channel state information of 
a client over an RB in each sub-frame (i.e., per eNB-UE antenna pair).
Each channel coefficient can be reliably represented using 6 magnitude bits and 6 phase bits, 
i.e. the channel quality indicator (CQI) and the channel phase indicator (CPI) each belongs to 
one of 64 levels (current LTE systems use a 4-bit CQI feedback, but we have observed that 
advanced functions like CoMP need a more precise representation in order to avoid precoding 
errors). 
Consequently, we need $12\times12$ uncompressed bits to represent the channel coefficients 
per RB per 1ms, which translates to a rate of 144Kbps for one RB and 14.4Mbps for a 20MHz 
LTE system (100 RBs). 
This is of course too high a requirement from a single small cell (note that a macrocell would
need to support several such small cells within each sector) and needs to be reduced in a way 
that does not affect the quality of control information.
\begin{figure}[htpb]
  \begin{center}
    \includegraphics[scale = 0.22]{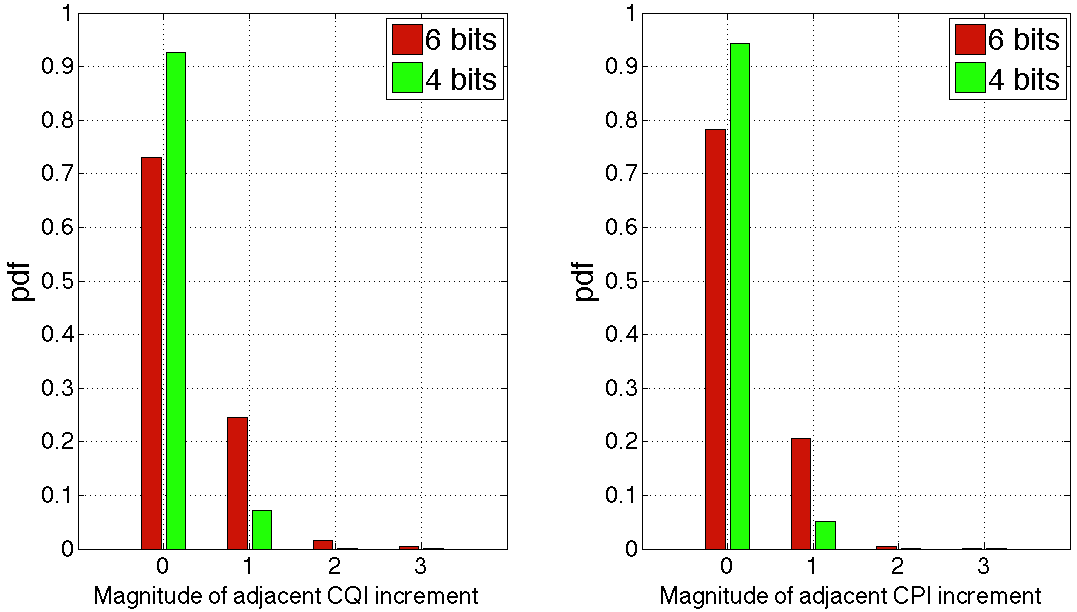}
    \caption{Probabilities of increments of CQI/CPI across consecutive sub-channels}
    \label{fig:freq_compr}
  \end{center}
\end{figure}

\begin{figure}[htpb]
  \begin{center}
    \includegraphics[scale = 0.22]{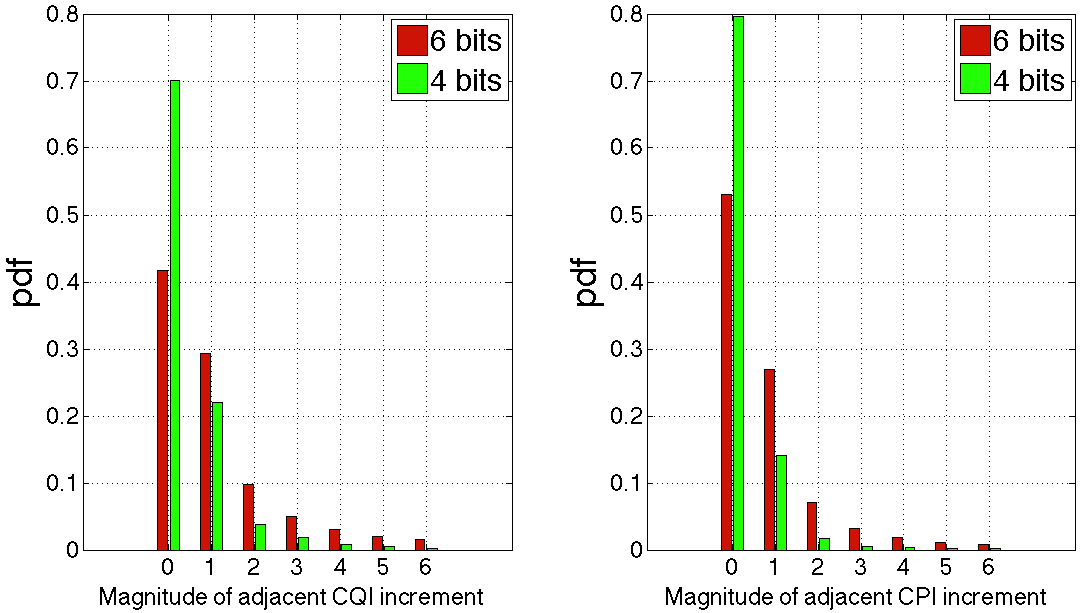}
    \caption{Probabilities of increments of CQI/CPI across consecutive sub-frames}
    \label{fig:time_compr}
  \end{center}
\end{figure}

However, as we show in Figure~\ref{fig:freq_compr}, CQIs and CPIs hardly ever change 
by more than one quantization level over one sub-channel.
In fact, with 6+6 bits of quantization, they remain the same with a 73\% probability and change 
by $\pm$1 level with a 24.5\% probability.
With such strong correlation, a simple sparsity transform such as the Fourier transform can
provide significant compression at almost no increase in distortion i.e., we can represent the 
channels across a resource block just by 6+6 bits of their mean (dc coefficient) which carries 
most of the information.
In addition, if we use the simple encoding scheme that we had described in Section~\ref{subsec:efficiency} 
for exploiting correlation across sub-frames, we would require 4.47+3.47 bits on average for 
every sub-frame after the first, using the probabilities computed in Figure~\ref{fig:time_compr}.

As a result, the total requirement over \name can be expected to be around 10Kbps per RB.
Since each small cell is expected to share the channels seen by each of its clients not just to
itself but also to at least 1-2 neighboring cells and assuming each small cell would need
coordination for $\sim50\%$ of its radio resources (the remaining resources would typically
be used with \texttt{IGNORE} for core users or \texttt{AVOID} i.e., eICIC, with the macro), the 
effective required rate is going to be $\sim3\times0.5\times10$ Kbps per RB.
This means that at the highest rate of coordination (i.e., per 1ms) with up to 2 neighbors for 
each client and for carrying full resolution channel state information without any increase in
distortion, \name would need to support about 15Kbps per RB, or equivalently 1.5Mbps over
20MHz.
The above analysis is for the average case scenario. 
In the worst case scenario, the control signaling rate would be a factor of two higher since 
that corresponds to using coordination over the entire set of resource blocks.

Due to its full duplex nature, having to send such control signaling traffic to
the macrocell from the small cell has almost no impact on the small cell's
capacity itself, we are getting this control signaling for free by transmitting
at the same time the small cell is receiving on the uplink frequency. 
However we do consume macrocell resources, whose impact we quantify next.

\subsubsection{Impact on macrocell capacity}
\begin{figure}[htpb]
  \begin{center}
    \includegraphics[scale = 0.23]{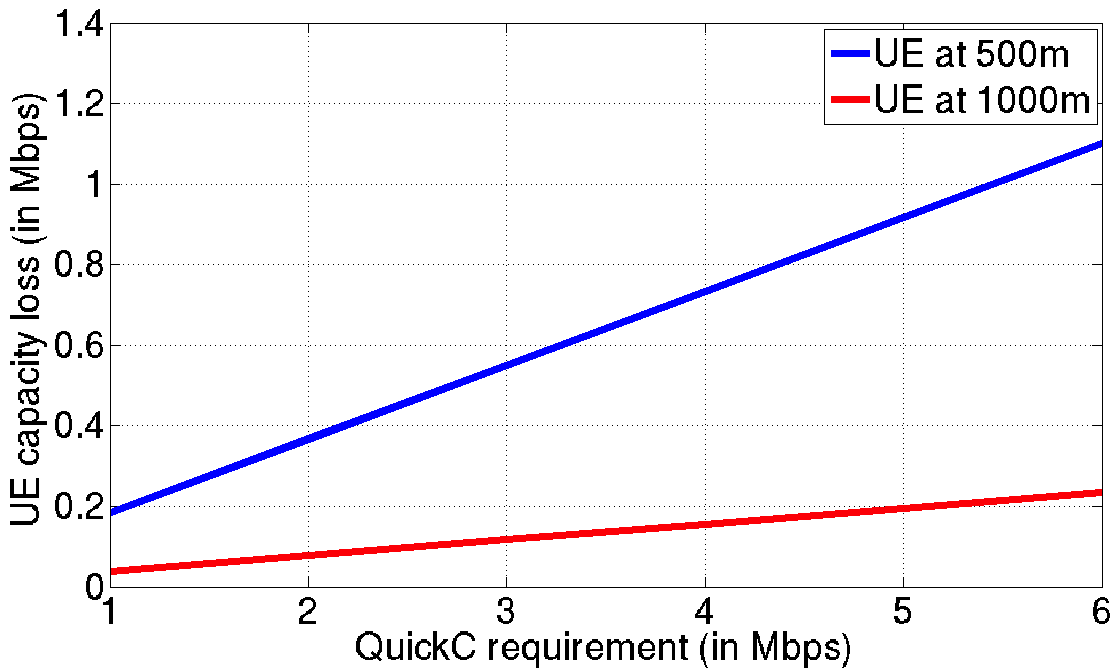}
    \caption{Average loss of uplink capacity for an UE connected to the macrocell
    		as a function of the \name capacity requirements.}
    \label{fig:impact_macro}
  \end{center}
\end{figure}

A single macrocell can be expected to cover 4 or more small cells within each sector,
so the total uplink capacity that it needs to provide for \name can be expected to be in
excess of 6Mbps.
Does that mean \name robs 6Mbps or more of uplink capacity straight from the UEs 
who could have otherwise used the resources in \name's absence?
The answer is not really, and the reason as we had mentioned in Section~\ref{subsec:efficiency} 
is that the small cell UEs typically see much stronger and static channels, and can very often 
operate at the highest modulation and coding rate even when deployed at the edge of the 
macrocell, while the corresponding rates over the same time-frequency resources are much 
lower for UEs located closer to the ground.
Figure~\ref{fig:impact_macro} shows the average \tit{loss} in the capacity of an UE connected
to the macrocell, and located at 500m and 1000m away, as a function of the capacity
requirements of \name. For e.g., even for an UE located in the core of the macrocell, 
providing 6Mbps on the uplink to \name is equivalent to denying only 1.1Mbps to the UE.
\label{subsec:impact_macro}

The main takeaway from this analysis is that although \name requires to carry control traffic on
the order of several hundreds of Kbps or even a few Mbps on the data channels of the macrocell,
the impact on the UE uplink throughput is much lesser.

\subsection{Interference management with \name}
In order to demonstrate the interference coordination benefits that \name can provide, we
simulate an urban hetnet consisting of a macrocell of 1km radius and a variable number
of microcells deployed within the coverage area of the macrocell.
The simulation parameters are summarized in Table~\ref{tab:simpar}.
We generate 1000 users over the region distributed uniformly at random, and randomly
assign them a velocity out of 1kmph, 5kmph and 30kmph (each velocity corresponds to
a different coherence time and therefore sees different rates of fading).
We implement a simple scheduler that works broadly as follows for each user.
\begin{Itemize}
	\item if the user is moving at 30kmph, it associates it with the macrocell irrespective of
	its proximity to microcells. Otherwise, 
	\item if the strongest channel seen by the user (in terms of Rx SNR) to any eNB is at 
	least 15 dB stronger than the second strongest, it associates the user with that eNB 
	who serves it using \texttt{IGNORE} (i.e., by treating interference as noise). Else,
	\item if the second strongest channel is at least 15dB better than the third strongest
	(if not, if the third strongest is 15dB better than the 4th strongest), it associates the user 
	with the two (or three) strongest eNBs who serve it using CoMP. Else,
	\item if there are four or more very strong interferers, one or more of them \texttt{AVOID}
	interference such that the rest can serve using CoMP.
	\item if there are multiple users associated to an eNB, it uses round-robin scheduling
	for serving the users.
\end{Itemize} 
Note that the above scheduling need not be the best way of managing interference in a
network and there might optimal schedulers that can provide higher network capacities; 
however, our focus here is solely on demonstrating the link-level gains that coordination 
techniques such as CoMP can provide when they have a low-latency control plane like
\name at their disposal.
\begin{table}[h]
	\centering
	\caption{Simulation Parameters}
	\begin{tabular}{|c|c|c|}
	\hline
   	\tbf{Parameter}		&	\tbf{Value/Setting}	\\ \hline
	Cellular Layout		&	1 macro (1km) \\
	(Radius)			&	Variable micro (200-600m)\\ \hline
	Channel Model		&	3GPP SCM~\cite{bib:3gppscm}\\ \hline
	Carrier Frequency	&	2GHz \\ \hline
	LOS				&	None \\ \hline
	Bandwidth		& 	5MHz (OFDM) \\ \hline
	Tx power (dBm)$^\dagger$	&	62, 30, 18$^*$ 	\\ \hline
	Rx noise floor (dBm)&	-105, -102, -98$^*$\\ \hline
	Height (m)		& 	32, 12.5, 1.5$^*$\\ \hline
	UE velocity (kmph)	& 	1, 5, 30 \\ \hline	
	\end{tabular}
    	\label{tab:simpar}
\end{table}
\squeezeup

{\small $^\dagger$includes antenna gains}
{\small $^*$macrocell, microcell, UE respectively}
\medskip

We use a MATLAB implementation~\cite{bib:matlab} of the 3GPP Spatial Channel Model 
(SCM)~\cite{bib:3gppscm} for generating the time-varying channel matrices. 
We use OFDM with 512 subcarriers over 5MHz, since the SCM documentation claims that
the model might not be suitable for systems with wider than 5MHz bandwidths.

\subsubsection{Impact on overall network capacity}
In order to evaluate the impact that a low-latency control plane like \name can have on
the overall capacity of a network, we simulate our test macrocell network with increasing
number of microcells (2-7) deployed per sector.
We compare the performance of a hetnet using the \name plane for coordination with
\begin{Enumerate}
	\item an \tit{ideal} network where each cell has its own piece of spectrum to operate 
	over without causing interference to anyone else (note that coordination techniques 
	like CoMP strive to make an interference-dominant network perform as if it were an 
	ideal interference-free network). The capacity of this network benchmarks the capacity
	scaling that is desired out of small cell deployments. 
	\item a hetnet using the X2 over IP control plane. We assume that such a control plane
	has a one-way latency of 10ms from a small cell to the packet core network, and
	hence a total latency of 20ms.
\end{Enumerate}

\begin{figure}[htpb]
  \begin{center}
    \includegraphics[scale = 0.22]{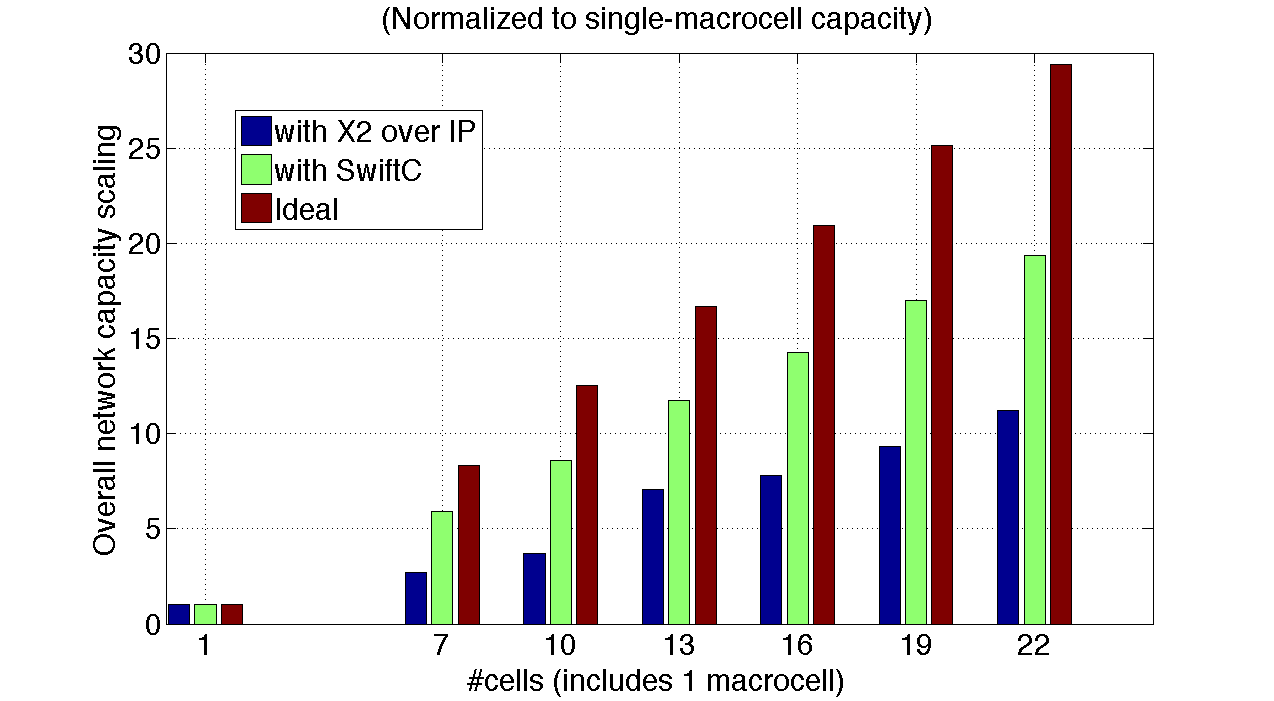}
    \caption{Overall network capacity in units of the capacity of a 1-macrocell network. With
    \name, the overall network capacity scales almost linearly with the number of small cells,
    while current control architecture based on X2 over IP fall way behind.}
    \label{fig:network_gains}
  \end{center}
\end{figure}

Figure~\ref{fig:network_gains} shows the overall network capacity for the three cases (Ideal, with
\name, with X2 over IP) normalized by the capacity of a single macrocell network.
The ideal capacity scales more than linearly (i.e. with N cells, the capacity is more than N times 
the capacity of a 1-cell network) due to the strong diversity gains in cell-splitting.
Although the network capacity with the \name plane does not provide the ideal-like diversity
gains, it does result in a nearly linear capacity scaling with density ($\sim19\times$ with 21 small cells), 
something that the network with X2 over IP clearly is not capable of (only 12$\times$ with 21 small cells). 
\squeezeup

\subsubsection{Impact on coordination gains}
\begin{figure}[htpb]
  \begin{center}
    \includegraphics[scale = 0.22]{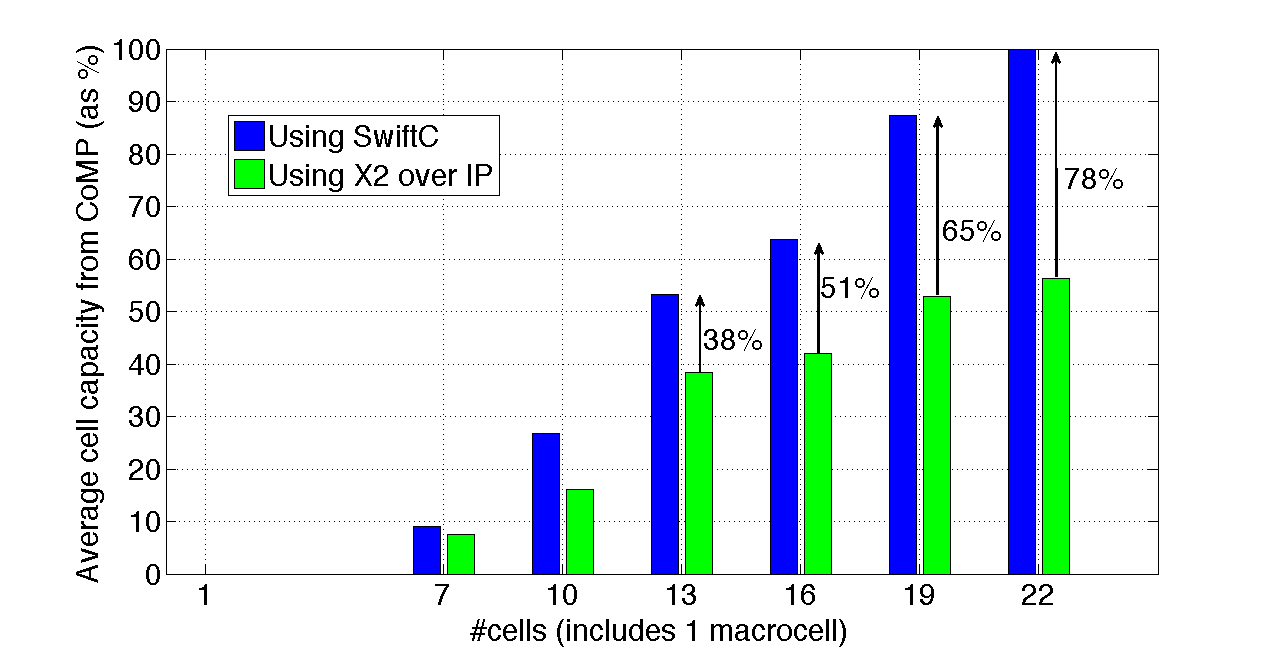}
    \caption{Relative capacities from CoMP transmissions using \name and X2 over IP.}
    \label{fig:comp_gains}
  \end{center}
\end{figure}

The total network capacity includes the macrocell capacity and a component of the small cell 
capacity that is delivered through techniques like \texttt{IGNORE} and \texttt{AVOID} which do 
not depend on the control plane quality.
In order to scrutinize the specific impact on coordination gains that \name can provide with 
respect to a standard X2 over IP control plane, we zoom into the component of network capacity 
that is delivered through CoMP transmissions.
Figure~\ref{fig:comp_gains} shows the relative gains from CoMP (as a \% of the plot maximum)
obtained via coordinating over (1) \name and (2) X2 over IP.
As the number of microcells deployed within a sector increases from 3 to 7, the CoMP gains using 
\name relative to X2 over IP increase from 38\% to 78\%.
Note that most of the CoMP transmissions happen for edge users who see strong interference
from one or more neighboring cells, so these results also imply that the \name plane can 
enable a significant increase in cell-edge throughput and thus effectively tackle the problem
of high interference in dense networks.
\squeezeup
\section{Conclusion}

\name shows that by physically decoupling the LTE data and control planes, and
by using LTE spectrum itself to implement the control plane, it is possible to
design and implement sophisticated network and interference management
strategies in dense LTE cellular networks. We are currently working on building
a testbed that embeds \name and evaluating the benefits with real world traffic.
We are also exploring how to integrate other network management functions such
as load management, handoffs etc into the \name control plane.

\end{document}